\documentclass[aps,pra,notitlepage,twocolumn]{revtex4-1}

\usepackage{amsmath}
\usepackage{amssymb}
\usepackage{fancyref}
\usepackage{tikz-qtree}
\usepackage[]{algorithm2e}
\usepackage{mdframed}
\usepackage{tikz}
\usepackage{graphicx}

\usetikzlibrary{shapes,snakes}
\definecolor{ltgray}{rgb}{0.95,0.95,0.95}
\definecolor{dkgray}{rgb}{0.33,0.33,0.33}
\usepackage{verbatim}
\newcommand{\id}{\mathbf{1}}
\newcommand{\fig}[1]{Fig.~\ref{#1}}
\DeclareMathOperator{\Tr}{Tr}

\usepackage{braket}
\begin{document}

\title{Local spin operators for fermion simulations}

 \author{James D. Whitfield}
 \email{JDWhitfield@gmail.com}
 \affiliation{Department of Physics and Astronomy, Dartmouth College, Hanover, New Hampshire 03755, USA}
 \affiliation{Department of Physics and Astronomy, University of Ghent, Krijgslaan 281 S9, B-9000 Ghent, Belgium}

\author{Vojt\v{e}ch Havl\'{i}\v{c}ek}
 \email{havlicek@itp.phys.ethz.ch} 
\affiliation{Institute for Theoretical Physics and Station Q Zurich, ETH Zurich, 8093 Zurich, Switzerland}

\author{Matthias Troyer}
\affiliation{Institute for Theoretical Physics and Station Q Zurich, ETH Zurich, 8093 Zurich, Switzerland}
\affiliation{Quantum Architectures and Computation Group, Microsoft Research, Redmond, WA 98052, USA}
\affiliation{Station Q, Microsoft Research, Santa Barbara, CA 93106-6105, USA}

\begin{abstract}
Digital quantum simulation of fermionic systems is important in the context of chemistry and physics. Simulating fermionic models on general purpose quantum computers requires imposing a fermionic algebra on spins. The previously studied Jordan-Wigner and Bravyi-Kitaev transformations are two techniques for accomplishing this task.  Here we re-examine an auxiliary fermion construction which maps fermionic operators to local operators on spins. The local simulation is performed by relaxing the requirement that the number of spins should match the number of fermionic modes. Instead, auxiliary modes are introduced to enable non-consecutive fermionic couplings to be simulated with constant low-rank tensor products on spins. We connect the auxiliary fermion construction to other topological models and give examples of the construction.
\end{abstract}

\maketitle
	
Quantum simulations have been a driver of quantum computing research since the earliest days of quantum computing~\cite{Feynman82}. 
In particular, electronic structure of interacting fermions is often highlighted as a prime application area~\cite{Georgescu14} with implications for condensed matter~\cite{Wecker15b,Kreula15} and quantum chemistry~\cite{Aspuru05,Babbush16}. Here, we contribute to this research direction with an alternative encoding of fermions in qubits.

Quantum computing is usually done using distinguishable two-level qubit systems. Thus, quantum computer simulations of fermions require imposing the fermionic statistics on the spin system through an encoding. Examples of such encodings include: the first quantized encoding~\cite{Zalka96,Kassal08}, unitary coupled cluster ansatze~\cite{McClean16,Wecker15}, the Bravyi-Kitaev~\cite{Seeley12,Tranter15},  and the Jordan-Wigner encodings~\cite{Somma02,Whitfield11}. The manipulation and extraction of physical features of a fermionic model then depends directly on the mapping employed.  

In this work, we study an auxiliary fermion encoding scheme for fermionic Hamiltonians that introduces additional degrees of freedom, but is manipulated with only local spin operators \cite{Verstraete05,Ball05,Farrelly14}. The previous works on the auxiliary fermion scheme~\cite{Verstraete05,Ball05,Farrelly14} are elaborated and generalized with the previous constructions recovered as specific cases of the framework. 

Of particular relevance to understanding the present work is the Jordan-Wigner mapping. The encoding \cite{Jordan28,Lieb61,Somma02,Whitfield11} maps one dimensional nearest-neighbor fermionic Hamiltonians to nearest-neighbor operators on spins. In higher dimensional fermionic systems however, manipulation of the fermionic state requires non-local spin operators after the Jordan-Wigner transform.  The cost of simulating non-local spin operators can be reduced using circuit optimization \cite{Hastings15} or teleportation techniques~\cite{Jones12}.  The closely related Bravyi-Kitaev encoding requires only logarithmically more local operators~\cite{Bravyi02,Seeley12,Tranter15}.  Both of these encodings require the same number of fermionic modes as the number of spins needed to encode the state, in contrast to the present method which requires increasing the number of qubits. Nevertheless, trading the increased dimension for reduction to a local spin Hamiltonian is still a desirable feature for quantum simulations.

Before giving the expression for the Jordan-Wigner transformation, consider the occupation representation of a fermionic state:
\begin{align}
\label{eq:occ}
\ket{n_0, n_1, n_2 \ldots n_{N-1} } &= \prod_{i=0}^{N-1}  \, (a^\dagger_i)^{n_i} \ket{\Omega}\,,
\end{align}
where $\ket{\Omega}$ is the fermionic vacuum state. The creation and annihilation operators satisfy the fermionic algebra:
\begin{align}
\left[a_i, a_j\right]_+ &= \left[a_i^\dagger, a_j^\dagger\right]_+ = 0 \,,& \left[a_i, a_j^\dagger\right]_+ &= \delta_{ij} \,.
\end{align} 
Consequently, the action of the annihilation operator is
\begin{align}
a_j & \ket{n_0, \ldots n_j, \ldots , n_{N-1}}  \nonumber\\ 
&= \delta_{n_j1} (-1)^{\Gamma_{j\mathbf{n}}} \ket{n_0,\ldots (n_j -1), \ldots n_{N-1}}  \,,
\end{align} 
with a non-trivial phase factor given by $\Gamma_{j\mathbf{n}}=\sum_{i=0}^{j-1}n_i$. 

The Jordan-Wigner encoding represents antisymmetric fermionic operators with tensor products of spin-$\frac12$ particle operators as:
\begin{align}
\label{eq:JW}
a_j  = \bigotimes^{j-1}_{i=0}Z_i \otimes A_j\,, 
\end{align} 
where $A_j=  \left( {X_j + iY_j} \right)/2$ is the single spin lowering operator acting on site $j$. The vacuum state is the +1 eigenstate of $\otimes_i^N Z_i$, $\ket{\Omega}=\ket{0...0}$.

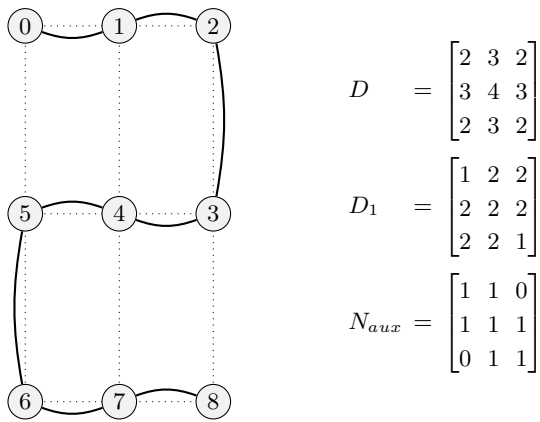
\begin{figure}[t]
\begin{tabular}{p{.5\columnwidth}r}

\setlength{\tabcolsep}{10pt}
\begin{tikzpicture}[scale=.5,baseline={([yshift={-20ex}]current bounding box.north)}]
    \draw[style=dotted] (0,0) grid[xstep=2.5cm, ystep=2.5*2cm] (5.0,2*5.0); 
    \foreach \x in {0,1,...,2}{                           
        \foreach \y in {0,1,...,2}{                       
        \pgfmathparse{int(3*(2-\y)+ \x)}\let \j\pgfmathresult
        \pgfmathparse{int(3*(3-\y)- \x-1)}\let \k\pgfmathresult
        \pgfmathparse{int(Mod(int(\y+1),2))}\let \b \pgfmathresult
        \ifthenelse{\b = 1}
        		{\node[fill=ltgray,circle,inner sep=2pt,draw](\j) at (2.5*\x,5*\y) { $\j$}}
    		{\node[fill=ltgray,circle,inner sep=2pt,draw](\k) at (2.5*\x,5*\y) { $\k$}}; 
        }
    }
    \draw [-, thick] (0) to[out=-20,in=200]   (1) (1) to[out=20,in=160] (2) (2) to[out=-80,in=80] (3) (3) to[out=200,in=340]  (4) (4) to[out=160,in=20]  (5) (5) to[out=260,in=100]  (6) (6) to[out=-20,in=200]  (7) (7) to[out=20,in=160] (8);
        \end{tikzpicture}
&
    \begin{tabular}{lcl}
    $D$&=&$\left[ \begin{matrix} 
     2& 3& 2\\
     3& 4& 3\\
     2& 3& 2\end{matrix}
     \right]$\\[5ex]
    $D_1$&=&$\left[\begin{matrix}
     1& 2& 2\\
     2& 2& 2\\
     2& 2& 1
    \end{matrix}\right]$\\[5ex]
    $N_{aux}$
    &=&$\left[\begin{matrix}
     1& 1& 0\\
     1& 1& 1\\
     0& 1& 1
    \end{matrix}\right]$\\
    \end{tabular}
\end{tabular}
\caption{The 2D Hubbard model provides an illustration of the advantage of the auxiliary fermion simulations over Jordan-Wigner and Bravyi-Kitaev transformed operators. On the left, the $L=3$ model is depicted with linear graph $G_1$ in curved bold face lines and with dotted lines indicating the Hubbard interaction graph, $G$.  On the right, $D=deg(G)$ is the degree of the modes in $G$, similarly $D_1=deg(G_1)$. Their difference gives the non-local degree, $D_{nl}$. This translates into the number of auxiliary fermions needed at each site following $N_{aux}=\textrm{ceil}(D_{nl}/2)$.}  
\label{fig:nonlocality2}
\end{figure}

The Jordan-Wigner transformation of the fermionic nearest-neighbor hopping term $a_p^\dagger a_q$ for $p < q$ is given by:
\begin{align} 
a_p^\dag a_q = A^\dagger_p \otimes \left(\bigotimes_{i=p+1}^{q-1} Z_i \right)\otimes A_q \,. 
\end{align}
This operator contains $|q-p|-1$ spin $Z$ operators in its spin representation. Thus, the spin Hamiltonian is local only when the fermionic Hamiltonian has consecutively ordered couplings. An exemplary case are one-dimensional nearest neighbor models, such as the one-dimensional Hubbard model. In general, however, the Hamiltonian will have non-local spin terms as illustrated by the 2D Hubbard model, \fig{fig:nonlocality2}.  

The explanation of the auxiliary fermion model begins with a derivation of the general form for the auxiliary couplings. We then analyze the spatial requirements of the model and illustrate the construction with two examples.  This is followed by discussion of state preparation and, lastly, we close the article with concluding remarks.

\paragraph*{Auxiliary coupling terms --}
The method for achieving locality in the Hamiltonian terms is to replace $h_{pq}^{old}=a_p^\dag a_q+ a_q^\dag a_p 
\mapsto h_{pq}=a_p^\dag\; M_{aux}{(pq)}\; a_q+ a_q^\dag\; {M_{aux}^\dag (pq)}\;a_p$  
in the fermionic model for sites $p$ and $q$ separated along the linear indexing.  The generalization to two-body four-point or higher interactions follows naturally: $a_p^\dag a_q a_r^\dag a_s\mapsto a_p^\dag M_{aux}(pq) a_{q} a_r^\dag M_{aux}(rs)a_s$.  Here, $M_{aux}(ij)$ is an operator on auxiliary fermionic modes introduced to cancel the JW non-local $Z$ chains without changing the physics of the original fermionic model. 
Besides mapping the Hamiltonian terms, this can be used for mapping correlation functions and other physically interesting observables.
As detailed below, the vacuum state $\ket{\Omega}$, is also modified to achieve the desired fermionic statistics.

Let us now turn to the required properties of $M_{aux}(ij)$.  We want any valid fermionic state of the original model in the new encoding to be stabilized by each of the $M_{aux}(ij)$.   
Moreover, we insist that $[M_{aux}, \, a_p] = 0$, implying that $a_p^\dag M_{aux} a_q\ket{\Omega}=a_p^\dag a_qM_{aux}\ket{\Omega}=a_p^\dag a_q\ket{\Omega}$.  We additionally require that the auxiliary couplings $M_{aux}$ mutually commute. Note that this condition is sufficient however it may not be necessary as we only need the vacuum state to be jointly stabilized.

To explore the algebraic nature of the model, consider the decomposition: $M_{aux}(pq)=i b_{p'} c_{q'}$.  Here the primed indices label auxiliary fermionic modes introduced adjacent to the corresponding mode of the original system. Previous examples in the literature~\cite{Verstraete05,Ball05,Farrelly14} relied on choices for $b_{j'}, \, c_{j'}$ of $(a_{j'}+a_{j'}^\dag)$ or $-i(a_{j'}-a_{j'}^\dag)$  which correspond to Majorana fermions.  However, we will show that this is more restrictive than necessary.  

In the same way that each $M_{aux}(pq)$ in the auxiliary fermion model mutually commutes, the terms in the toric model~\cite{Kitaev97,Pachos12} also mutually commute.  In both models, a code space defined by the joint eigenspace of the mutually commuting operators is preserved. In the present model, however, the excitation space is irrelevant to the construction.  Also note that the auxiliary fermion model supports both odd and even particle number sectors.  This can be understood by noting the first site on the index backbone never participates in any of the non-local couplings. This allows a single fermion to be inserted or removed without modifying the code space.  One can think of an analogy using the toric code whereby a single excitation (rather than a pair string-like excitations)  is allowed by appropriate modification of the boundary terms.

In general,  
we can represented algebraic operators on spins by:
\begin{align*}
  b_{p'} &= \bigotimes_{i=0}^{p'-1}Z_i \otimes B_{p'} 
& c_{q'} &= \bigotimes_{i=0}^{q'-1}Z_i \otimes C_{q'} \,,
\end{align*}
with $B_{p'}$ and $C_{q'}$  single spin operators. Any non-trivial Bogoliubov transformation of the following form suffices: 
\begin{align}
b_m&= \left( \alpha^{-1} a_m + \alpha \, a_m^\dag \right)  \\
&=  \bigotimes_{i=0}^{m-1}Z_{i}\otimes \frac{1}{2}\left[ (\alpha^{-1}+\alpha) X  + i(\alpha^{-1}-\alpha) Y \right]_m\\
 &=  \bigotimes_{i=0}^{m-1}Z_{i}\otimes B_m
\end{align}
with $\alpha=e^{-i\theta}$ for all real $\theta$.
The choice of $\alpha = 1$ or $\alpha = i$, corresponds to the Majorana fermions with $B=X$ and $B=Y$ respectively. 
The eigenstates of $B$ are: 
\begin{equation}
\ket{\pm_b}=\frac{\ket{0}\pm \alpha \ket{1}}{\sqrt{2}}
\label{eq:eigvecs}
\end{equation} 
with eigenvalues $\pm 1$. 

The fermionic phase factor gives rise to the spin-locality of the Jordan-Wigner transform.  In the occupation number basis, Eq.~\eqref{eq:occ}, the modes are in a fixed ordering. All other orderings of the modes map back to this state with either a $+1$ or $-1$ phase factor.  The Jordan-Wigner $Z$ chains compute this reordering factor.  In this model, the reordering factor is computed using the code space of $\{M_{aux}(pq)\}$. Consider the action of $\hat K=\prod_i^N a_i^\dag$ on the vacuum state. With the JW representation \eqref{eq:JW} of $\hat K$, the auxiliary mode $j'$ will store parity $p_j=\sum^M_{k=j+1} n_k$. This follows from Eq.~\eqref{eq:eigvecs} since $Z\ket{\pm}=\ket{\mp}$. Thus, the information about the parity is stored locally in the correlation of the $i'$ and $j'$ spins with values of $s_i,s_j\in\{+,-\}$. The product $s_is_j$ gives the phase factor associated with the reordering needed to implement $a_i^\dag a_j+a_j^\dag a_i$.

\paragraph*{Spatial requirements --}

The number of additional auxiliary fermionic modes will depend on how the interaction graph differs from a linear graph.  In the linear graph all fermionic modes have only two neighbors.  Whenever a fermion participates in non-local interactions involving $D>2$ other fermionic modes, then the present model requires $\textrm{ceiling}(D/2)$ auxiliary fermions to be introduced for completely local simulation.  As the number of auxiliary fermionic modes increases, the operator locality also must increase. The creation or annihilation of a fermion at a site requires that all auxiliary modes affiliated with that site also be updated. Next, we show that each auxiliary mode can couple up to two non-local neighbors.

We let $B_\perp$ be the orthogonal partner to $Z$ and $B$, such that $\Tr{(ZB)} = \Tr{(ZB_\perp)} = \Tr{(B B_\perp)} = 0$. It follows from anticommutation that:
$\pm iBZ = B_\perp$. Since $B$ and $Z$ have only one mutually orthogonal partner within $\mathfrak{su}(2)$, only up to two non-local couplings can be connected to a single auxiliary mode.  

We divide the analysis into two cases (assume that $p<q<r$): first, with $M_{aux}(pq)$ and $M_{aux}(pr)$ and second with $M_{aux}(pq)$ and $M_{aux}(qr)$.  These are the only two relevant cases because the indexing is linear.   
Then $M_{aux}(pq)$ is given by: 
\begin{equation}
\begin{array}{c@{\hskip1.5em}|cccccc}
   		M_{aux}(pq) &p &p'      &  & q & q'   & \\
		\hline
ib_{p'} & Z & iB_\perp   & &  & & \\
c_{q'} &  Z & Z  & Z \ldots Z &Z & C\\
\hline
ib_{p'} c_{q'} & \id & B & Z \ldots Z & Z  & C 
\end{array}
\end{equation}
In the first case, the auxiliary couplings
share a common node $p$. One can therefore write the spin operators as:  
\begin{equation}
\label{tab:diagram2}
\begin{array}{c@{\hskip1.5em}|cccccc}
   		&p'  &        & q' &     & r' & \\
		\hline
M_{aux}(pq)& B &   Z...Z& C & & &\\
M_{aux}(pr)& B' &   Z...Z& Z & Z...Z& C'\\
\end{array}
\end{equation}
The operators $B', C'$ can differ from $B,\,C$. 
Since $C$ and $Z$ anti-commute, for $M_{aux}(pq)$ and $M_{aux}(pr)$ to commute, $B$ and $B'$ must also anti-commute.  Since $\Tr\left([ B, B' ]_+ \right) = 2 \Tr(B B') = 0$, it follows that $B'$ must be orthogonal to $B$ in the Hilbert-Schmidt norm.  However, we have already shown that $(B_{\perp})_\perp = \pm B$, and therefore $B' = \pm B_\perp$. 

In the second case, where the non-local links share a common site $q$, the operator acting on $q$ must be the same for both of the auxiliary couplings since the two couplings will only overlap at $q$.

\paragraph*{Examples -- }

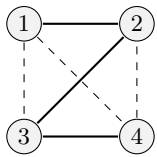
\begin{figure}[t]
\begin{tikzpicture}[scale=.5]
\node [fill=ltgray,circle,inner sep=2pt,draw] at (.5,.5) {3};
\draw[-,thick] (1, 0.5) -- (3,0.5);
\draw[-,dashed] (.85355, 3.14645) -- (3.1465,0.85355);
\node [fill=ltgray,circle,inner sep=2pt,draw] at (3.5,.5) {4};
\draw[-,dashed] (3.5,1) -- (3.5,3);
\node [fill=ltgray,circle,inner sep=2pt,draw] at (3.5,3.5) {2};
\draw[-,thick](3.14645,3.14645) -- (.85355,.85355);
\draw[-,thick] (3,3.5) -- (1,3.5);
\node [fill=ltgray,circle,inner sep=2pt,draw] at (.5,3.5) {1};
\draw[-,dashed] (.5,3) -- (.5,1);
\end{tikzpicture}
\caption{The $K_4$ graph is the completely connected graph with four sites.  The solid lines indicate the linear indexing and the dashed edges indicate non-local couplings.}
\label{fig:k4}
\end{figure}

As a simple example, let us consider the simulation of the completely connected four mode fermionic system depicted in \fig{fig:k4}. First, we must choose a basis for the $M_{aux}$ operators, construct the invariant vacuum state, finally, we can give expressions for one-body and two-body couplings. 

Without loss of generality, we fix the $\alpha$-gauge for the auxiliary couplings such that,
\begin{align}
\begin{array}{c@{\hskip1em}ccccccccc}
& 1 &1' & 2 & 2' & 3 & 3' & 4 & 4'\\
\hline
M_{13}& &X &Z&Z & Z & X &\\
M_{14}& &Y &Z&Z & Z & Z & Z &Y \\
M_{24}& &  & &X & Z & Z & Z &X \\
\hline
\end{array}
\label{eq:M}
\end{align}
Note that the choice of $M_{13}$ and $M_{14}$ fix $M_{24}$ by the commutation requirements.

Non-local one-body terms for the $K_4$ graph follow from \eqref{eq:M} and \eqref{eq:JW}.
\begin{align}
\begin{array}{c@{\hskip2em}ccccccccc}
				  & 1      &1' 	& 2 & 2' & 3 & 3' & 4 & 4'\\
				  \hline
a_1^\dag M_{13}a_3& A^\dag & -iY&   &  & A & X &\\[1.5ex]
a_2^\dag M_{24}a_4&        &    &  A^\dag &-iY  &   &   & A &X \\[1.5ex]
a_1^\dag M_{14}a_4& A^\dag &  iX&     &  &   &   & A &Y \\
\hline
\end{array}
\end{align}

The remaining local hopping terms are of the form $a_p^\dag a_{p+1}=A^\dag_p \otimes Z_{p'} \otimes A_{p+1}$. The one-local number operator is given by $a_k^\dag a_k=(\id-Z_k)/2$.  

The two-body terms of this example are easily obtained as well. The two-body terms consisting of two-point interactions, e.g. $a_i^\dag a_j^\dag a_j a_i$, are straight-forward products of the spin representations of $a_i^\dag a_i$ and $a_j^\dag a_j$.  Similarly, three-point interactions, e.g. $a_i^\dag a_k a_j^\dag a_j$, are also product of the spin-representation of $a_j^\dag a_j$ and $a_i^\dag M_{aux}(ik)a_k$, when $i$ and $k$ have non-consecutive indices.  The four-point interactions do not require the auxiliary coupling due to anti-commutation relations.  

Consider, for example, the term $a_1^\dag a_4^\dag a_2 a_3$ which occurs in the quantum simulation of molecular hydrogen using a minimal basis \cite{Whitfield11}.  Here, we can avoid the use of auxiliary spins by rearranging the term as $a_1^\dag a_2 a_4^\dag a_3$.  Now the term is a product of linearly-local hopping terms; each of which can be simulated without appeal to the auxiliary couplings.  This points out the importance of exercising the commutation relations to minimize the tensor weight of the simulated term.  

While this example illustrates the model, it is not chosen to highlight the decisive advantages of the scheme.  In fact, the Jordan-Wigner and Bravyi-Kitaev Hamiltonian on four sites also only has fourth order tensor products, but requires half as many spins.  When more non-local couplings are present the auxiliary fermion model 
will offer decisive advantages as 
illustrated with the next example. 

The second example is square 
lattices with $N=L^d$ in $d=2$ and $d=3$ dimensions. In our analysis, we only 
consider $L\rightarrow \infty$ bulk terms but see \fig{fig:nonlocality2} for $L=3$ example.  Each 
bulk site has $2d$ neighbors.  Subtracting the linear degree of $2$, each site
participates in $2d-2$ non-local interactions. By the arguments given earlier,
$d-1$ auxiliary modes are needed for each mode in the bulk. The maximum tensor 
product needed for the local simulation of interaction term involves $2d-2$ 
spin operators. Note that this is independent of $L$ as the simulation now only
depends on the local properties of the interaction graph.

\paragraph*{State preparation --}

The model is closely connected to topological models found in error correction codes~\cite{Pachos12}. To have robust error correction,
topological structures are used to store information as non-locally as possible. However, here we are attempting to store information strictly locally.

To highlight this overlap, a simple expression for the projected vacuum state can be borrowed from topologically non-trivial models~\cite{Dennis02}: 
\begin{equation}
\ket{\Omega} = \prod_{(pq)\in E} \frac{\id+M_{aux}(pq)}{\sqrt{2}}\ket{0..0} \,.
\label{eq:vac}
\end{equation} 
Because $M_{aux}(pq)^2=\id$, this is a projective operation. We begin by creating the state $\ket{00..0}$ and proceed to projectively measure each auxiliary coupling.  
If the measurement outcome for an auxiliary coupling, say $M_{aux}(pq)$, is $-1$, then changing its sign is a matter of applying $Z_{p'}$ or applying $Z_{q'}$ to the measured state.  When the mode participates in two non-local couplings, the error will propagate to the other non-local coupling.  Therefore, it is simplest if an auxiliary mode 
with only one non-local nearest-neighbor is chosen.  Otherwise, one should follow the linked chain of non-local modes applying $Z$ operator at each endpoint until the linked chain ends.  This is possible so long as no closed loops of auxiliary couplings are present. 

Note that the no closed loop restriction is not a serious limitation. For any closed loop of non-local couplings, e.g. $\{M_{aux}(pq), \, M_{aux}(qr),... \, ,M_{aux}(sp)\}$, we can take $M'_{aux}(p+1,q)$ instead of $M_{aux}(pq)$. The coupling from $p$ to $p+1$ can be done locally allowing nearly the same connectivity to be achieved. Moreover, we may now ignore non-trivial Wilson loops as done in~\cite{Ball05} since there are no loops by construction.

It is interesting to note that both the vacuum state in \label{eq:vac} and the completely filled states are  invariant under $M_{ij}$. Note that this must be true as $b_m^\dag=a_m$ is the hole creation operator with respect to the filled vacuum state.  Hence, the action of $\{b_m^\dag,b_k\}$ must also be antisymmetric.

Next, consider the preparation of states with $N$ fermions.  This is accomplished most straightforwardly by applying the Jordan-Wigner representation of $\hat K=\prod_i^N a_i^\dag$ on the vacuum state.  The action on the state can be simplified when the structures of the desired state and auxiliary lattice is known beforehand.  In this case, the occupied modes can be acted on with $X$. The auxiliary mode $k'$ is acted upon with $Z^p$ where $p$ is the parity of occupied modes to the right of $k'$.

\paragraph*{Conclusion -- }
In this work, we have studied the auxiliary fermion scheme for encoding fermionic states which enables highly localized manipulation. This encoding is designed to make information as accessible as possible.  This makes the system easier for both experimentalist and noise sources to modify the information. While the Jordan-Wigner encoding requires non-local manipulation of the state, recent numerical analysis suggest that it is more robust against noise than the Bravyi-Kitaev encoding~\cite{Sawaya16}. This gives some support for the hypothesis that local fermionic encodings will be less robust against noise sources.  Future work will investigate this trade-offs between noise and encoding for the locally encoded model described in this paper.  Other directions of this work is the comparison against related ideas for reducing tensor product rank of fermionic simulations~\cite{Bravyi02,Subasi16,Reiner16} and applications to adiabatic computation~\cite{Biamonte11,Babbush14}.  

This work was supported through the ERC grants QUTE and SIMCOFE and by the Swiss National Science Foundation through NCCR QSIT. We thank Frank Verstraete, Lei Wang, Ye-Hua Liu and Alexey Soluyanov for useful discussions and Peter Love for comments on the manuscript. 

\bibliography{local_fermionic_spin_half_algebra}

\end{document}